\documentclass[12pt]{article}
\usepackage{times}
\usepackage{geometry}
\usepackage[utf8]{inputenc}
\usepackage{enumitem,amssymb}
\usepackage{ragged2e}
\usepackage[dvipsnames]{xcolor}
\usepackage{graphicx}
\newlist{thematic}{itemize}{8}
\setlist[thematic]{label=$\square$}
\usepackage{pifont}

\usepackage{wrapfig}

\begin{document}
\begin{flushleft}
\huge
Astro2020 Science White Paper \linebreak

Simultaneous Measurements of Star Formation and Supermassive Black Hole Growth in Galaxies\linebreak
\normalsize

\noindent \textbf{Thematic Areas:} \hspace*{60pt} 
  $\boxtimes$    Galaxy Evolution   \linebreak
  
\textbf{Principal Author:}

Name: Alexandra Pope
 \linebreak						
Institution: University of Massachusetts Amherst	
 \linebreak
Email: pope@astro.umass.edu	
 \linebreak
Phone:  413-545-1769
 \linebreak
 
\textbf{Co-authors:} 
\small{
Lee Armus (IPAC/Caltech), 
Eric Murphy (NRAO), 
Susanne Aalto (Chalmers University of Technology), 
David Alexander (Durham University),
Philip Appleton (IPAC/Caltech),
Amy Barger (University of Wisconsin Madison),
Matt Bradford (JPL),
Peter Capak (IPAC/Caltech),
Caitlin Casey (University of Texas Austin), 
Vassilis Charmandaris (University of Crete),
Ranga Chary (IPAC/Caltech),
Asantha Cooray (University of California Irvine),
Jim Condon (NRAO)
Tanio Diaz Santos (Universidad Diego Portales),
Mark Dickinson (NOAO),
Duncan Farrah (University of Hawaii),
Carl Ferkinhoff (Winona State University),
Norman Grogin (STScI),
Ryan Hickox (Dartmouth College), 
Allison Kirkpatrick (University of Kansas),
Kohno Kotaro (University of Tokyo), 
Allison Matthews (University of Virginia),
Desika Narayanan (University of Florida),
Dominik Riechers (Cornell University),
Anna Sajina (Tufts University),
Mark Sargent (University of Sussex),
Douglas Scott (University of British Columbia),
J.D.~Smith (University of Toledo),
Gordon Stacey (Cornell University), 
Sylvain Veilleux (University of Maryland),
Joaquin Vieira (University of Illinois)}
 \linebreak
\end{flushleft}
\vspace{-0.2in}
\noindent \textbf{Abstract:} Galaxies grow their supermassive black holes in concert with their stars, although the relationship between these major galactic components is poorly understood. 
Observations of the cosmic growth of stars and black holes in galaxies suffer from disjoint samples and the strong effects of dust attenuation. 
The thermal infrared holds incredible potential for simultaneously measuring both the star formation and black hole accretion rates in large samples of galaxies covering a wide range of physical conditions.
{\it Spitzer} demonstrated this potential at low redshift, and by observing some of the most luminous galaxies at $z\sim2$. {\it JWST} will apply these methods to normal galaxies at these epochs, but will not be able to generate large spectroscopic samples or access the thermal infrared at high-redshift. An order of magnitude gap in our wavelength coverage will persist between {\it JWST} and ALMA.
A large, cold infrared telescope can fill this gap to determine when (in cosmic time), and where (within the cosmic web), stars and black holes co-evolve, by measuring these processes simultaneously in statistically complete and unbiased samples of galaxies to $z>8$. A next-generation radio interferometer will have the resolution and sensitivity to measure star-formation and nuclear accretion in even the dustiest galaxies. Together, the thermal infrared and radio can uniquely determine how stars and supermassive blackholes co-evolve in galaxies over cosmic time. 

\pagebreak

\section*{Co-evolution of galactic star formation and black hole growth}
\vspace{-0.1in}

\setlength{\parindent}{5ex}

One of the most important advances in the study of galaxies over the last two decades has been the discovery that the masses of the central supermassive black holes (SMBHs) and the stars are correlated in the local Universe, with a stellar bulge to SMBH ratio of about 700:1 (Kormendy \& Ho 2013).  Given the vastly different physical scales between the central supermassive black hole and the stellar bulge in a galaxy, it is remarkable that the two can influence each other through different stages of galactic growth (Alexander \& Hickox 2012).
By studying the relative black hole accretion rates (BHARs) and star formation rates (SFRs) in large samples of galaxies as a function of redshift and environment, we can rewind the clock on the lifecycle of galaxies, learning how, and when, they built up their stars and supermassive black holes. 

\begin{wrapfigure}{r}{0.6\textwidth}
  \begin{center}
  \vspace{-0.45in}
\includegraphics[width=10 cm]{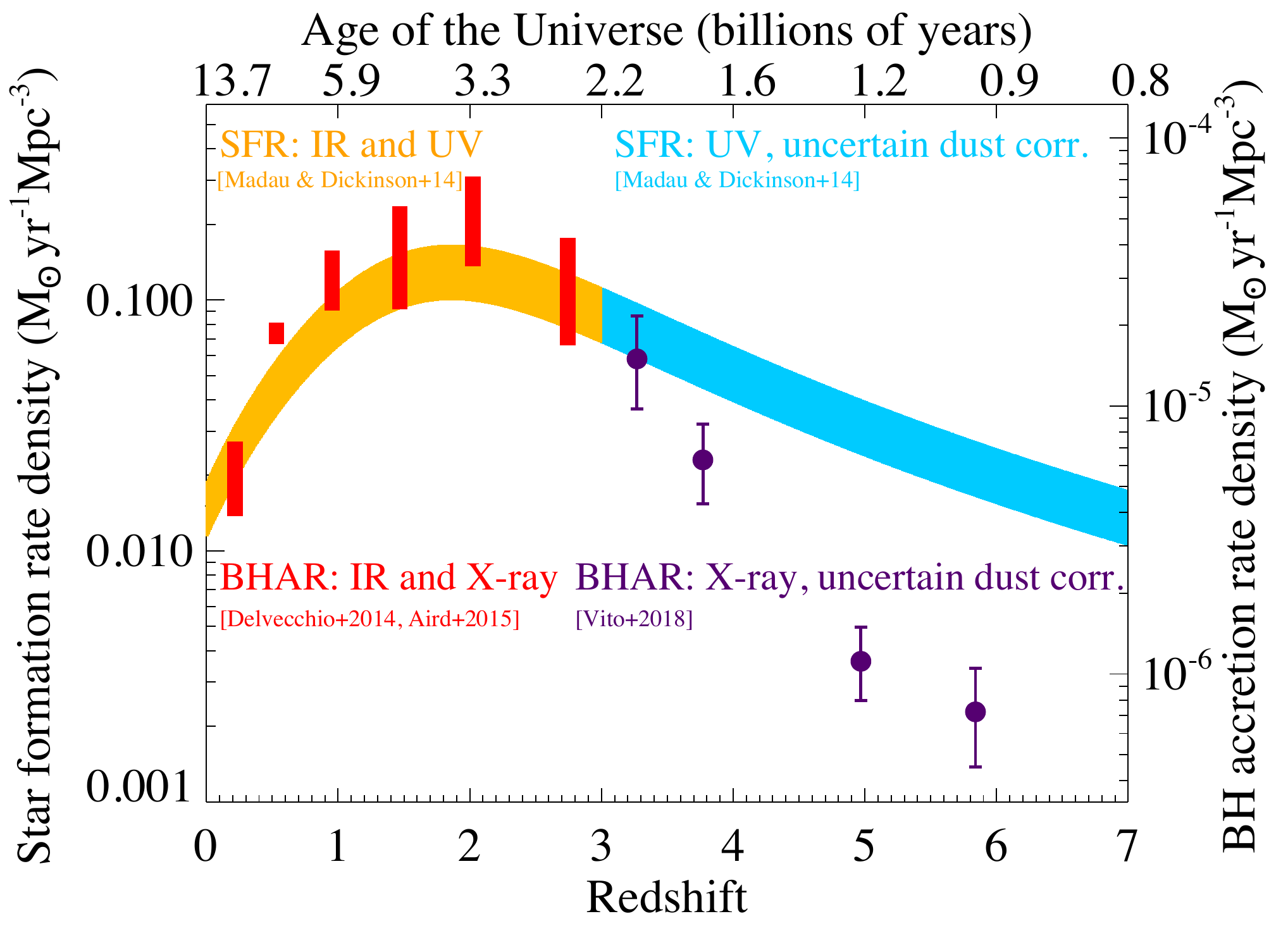}
  \end{center}
  \vspace{-0.32in}
  \caption{{\small 
  Relative evolution of the cosmic SFRD (shaded curves, left $y$-axis) and the BHARD (data points, right $y$-axis) from different wavelength surveys: at $z\,{>}\,3$ both are severely incomplete due to dust. 
  }
\label{fig:sfrd}}
  \vspace{-0.1in}
\end{wrapfigure}
The volume-averaged history of star formation (star formation rate density, SFRD) and black hole growth (BHAR density, BHARD) in the Universe both appear to increase up to a peak period at $z\,{=}\,1$--3 (referred to as ``Cosmic Noon", the epoch when half of the present-day stellar mass was produced), and then decline at higher redshifts (Figure 1). At these high redshifts, the growth rate of SMBH's may decline more steeply than star formation, indicating changes in their relative growth in the very early Universe. However, these constraints on the average SFRD and BHARD have been based on studies of different galaxy populations at different wavelengths. 
Piecing together these studies to produce a global picture of the evolution of galaxies as a function of lookback time is fraught with uncertainties as it masks the underlying diversity in the populations, and has observational biases.
The average SFRD is calculated from galaxies selected in ultra-violet (UV), optical or infrared imaging surveys that rely on photometric redshifts and broad-band estimates of the SFR and corrections for the significant dust attenuation. On the other hand, the average BHARD is typically estimated from galaxies selected from deep X-ray surveys, then extrapolated to take into account the significant, undetected, Compton-thick population (Hickox \& Alexander 2018).
{\bf In order to understand the build-up of stars and supermassive black holes over cosmic time---a critical missing ingredient for a self-consistent theory of galaxy evolution---we need measurements of the SFRs and the BHARs in the \underline{exact same} galaxies over the last 12 billion years.}

Most of the light from star formation and black hole growth over cosmic time is obscured by dust and re-radiated in the thermal infrared ($\sim\,$5--500$\,\mu$m). Current observations of this obscured activity is highly incomplete at $z\,{>}\,3$ (Madau \& Dickinson 2014 and references therein, Figure 1). 
Unfortunately, deep single-dish far-infrared observations of distant galaxies have thus far been limited by sensitivity and source confusion, while existing interferometric radio and sub-mm facilities lack the capabilities to simultaneously map wide ($>$ several deg$^2$) areas and reach the necessary depths to assemble statistical samples. 
UV- and optical-based estimators of the SFR in galaxies have been employed well into the epoch of reionization, but they are dependent upon uncertain, and often quite large, corrections accounting for the fraction of ionizing photons absorbed by dust. Similarly, X-ray surveys have detected strong AGN back to $z>6$, but miss the dominant population of obscured AGN and those with very strong contributions from star formation. The radio and thermal infrared wavelength regimes are not limited by these powerful biases and, together, they offer great potential for measuring the obscured star formation and black hole growth to definitively answer the following questions: {\bf How do stars and supermassive blackholes co-evolve in galaxies over cosmic time?} {\bf What are the dust-enshrouded SFRs and BHARs at $z>3$ ?} {\bf What drives feedback as a function of mass and environment and helps to regulate star formation at $z<3$?}

\vspace{-0.2in}

\section*{Infrared Signatures of Star Formation and Black Hole Accretion}
\vspace{-0.1in}

\begin{wrapfigure}{r}{0.55\textwidth}
  \begin{center}
  \vspace{-0.45in}
\includegraphics[width=9cm]{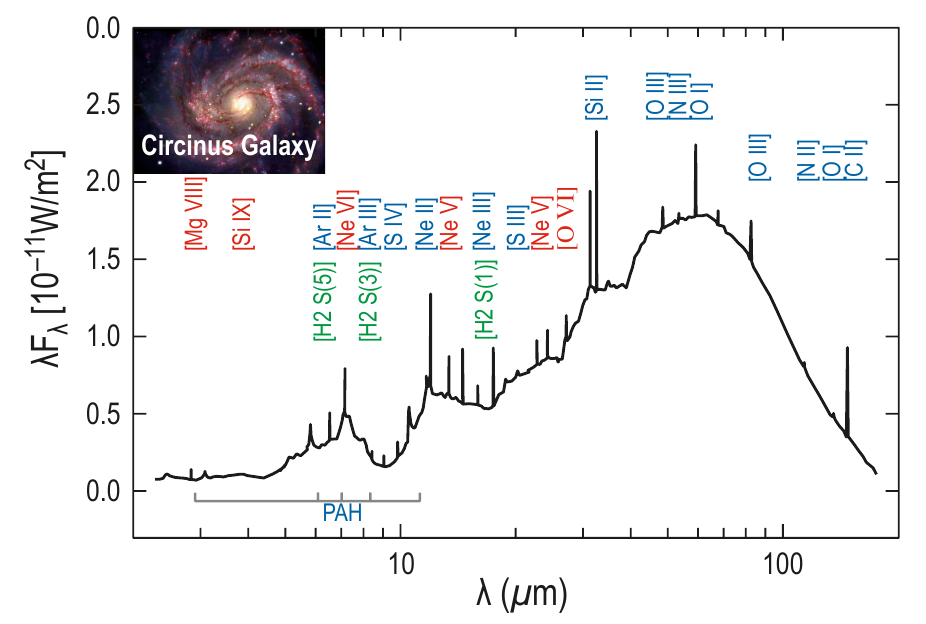}
  \end{center}
  \vspace{-0.2in}
  \caption{{\small 
 Circinus Galaxy infrared spectral energy distribution highlighting key spectral tracers of star formation (blue), black hole accretion (red) and warm molecular gas (green). Data from ISO (Genzel \& Cesarsky 2000). 
  }
\label{fig:sed_ir}}
\end{wrapfigure} 
The thermal infrared contains direct and indirect tracers of the SFR, the AGN luminosity and hence the BHAR, and the black hole mass in galaxies (Figure 2). By employing multiple independent tracers, one can mitigate uncertainties that plague each indicator, enabling more accurate estimates of all of these key quantities to be determined for large samples of individual galaxies. Here we describe these tracers in detail.\\
\vspace{-0.1in}

\noindent{\bf Dust Spectral Features} Polycyclic aromatic hydrocarbon (PAH) features are emitted by small dust grains stochastically heated by UV photons.  These grains are easily destroyed in the hard radiation fields surrounding AGN, but, in starburst galaxies, they are among the strongest features in the infrared (at 6.2, 7.7, 8.6, 11.3 and 17$\mu$m), easily reaching a few percent of the bolometric luminosity. This makes them ideal for determining both the redshifts and the SFRs of distant galaxies. {\it Spitzer}/IRS classified hundreds of galaxies from $z\,{=}\,0$--4 using these infrared spectral diagnostic features, albeit limited to the most luminous galaxies at high redshift (e.g.~Yan et al.~2005; Houck et al.~2005; Brandl et al.~2006; Smith et al.~2007; Armus et al.~2007; Pope et al.~2008; Riechers et al.~2014).

Obscured AGN are surrounded by dust which gets heated to high temperatures, and they therefore have a much warmer dust spectrum than starburst galaxies, with excess emission around 5--$10\,\mu$m (even the warmest starburst galaxies typically peak at much longer wavelengths, 50--$75\,\mu$m rest-frame). The equivalent width (EQW) of the PAH features thus provides a ratio of starburst-to-AGN power. 
The PAH EQW is often used in concert with the atomic fine-structure line ratios in a diagnostic diagram to identify the power sources in very dusty galaxies (e.g., Genzel et al., 1998; Laurent et al.~2000; Armus et al., 2007; Veilleux et al.~2009; Petric et al.~2011). \\
\vspace{-0.1in}

\noindent{\bf Atomic Fine-Structure and H$_2$ Emission Lines} Atomic fine-structure emission lines, such as [NeII] 12.7, [NeIII] 15.5, and [SIII] 18.7, 33.5 are produced in the H{\sc ii}-regions surrounding young stars and have been shown to correlate extremely well with the star-formation rates derived from the dust continuum and hydrogen recombination lines (e.g., Ho \& Keto, 2007; Inami et al., 2013). 
Features such as [SIV] 10.5, [NeV] 14.3, 24.3, and [OIV] 25.9 are tracers of highly ionized gas (35--97 eV). 
While such lines can be produced around very hot stars (Bernard-Salas et al., 2001; Pottasch et al., 2001; Smith et al., 2004; Devost et al., 2007), they are unusually strong in AGN, and can be effective tracers of the AGN power (e.g., Genzel et al., 1998; Lutz et al., 2003; Armus et al., 2006, 2007; Veilleux et al.~2009). The lines can also be used to estimate to AGN luminosity and mass accretion rate (e.g, Gruppioni et al. 2016).
Furthermore, in these AGN-dominated systems, the widths of the mid-infrared [NeV] and [OIV] emission lines have been shown to correlate with the mass of the central supermassive black hole in local Seyfert galaxies, as determined by reverberation mapping techniques (e.g., Dasyra et al. 2008). 

Fine-structure lines of highly depleted elements (e.g. Fe, Si) can also be used to pinpoint gas which is shock heated. Together with enhanced H$_2$ emission lines in the mid-infrared from warm (200--1000K) gas (see White Paper by Appleton et al.), they can be used to find and characterize galaxies experiencing widespread SNe and/or AGN driven feedback over a large range in galaxy mass. Feedback is implicated as a key ingredient in the co-evolution of galaxies and supermassive black holes, but has yet to be studied in statistical samples for which the concurrent SFRs and BHARs are also measured for all galaxies. \\
\vspace{-0.1in}

\noindent{\bf Dust Continuum} The shape of the mid-to-far IR dust continuum itself is sensitive to the relative heating by AGN and star formation (e.g., Veilleux et al.~2009; Kirkpatrick et al., 2015). With measured redshifts, the mid-IR and far-IR continuum emission can be fit with starburst and AGN templates to constrain the relative fraction of the bolometric luminosity (best measured in the infrared where most of the reradiated power emerges) heated by young stars and accreting supermassive black holes. For each galaxy, simple conversion between the far-infrared luminosity and the emitted flux from young stars (e.g., Murphy et al., 2011; Kennicutt et al., 1998) allows for a derivation of the current SFR in each source. The continuum at mid-IR wavelengths provides an estimate of the AGN luminosity in systems where it dominates. Our current census of the dust-obscured SFRD to cosmic noon (e.g., Madau \& Dickinson, 2014; Casey et al. 2018) is based almost entirely on these IR-continuum-based SFRs. 

These continuum-based estimates of the ongoing star formation in galaxies can be done quickly for thousands or millions of galaxies in large area surveys, but they always rely on model fitting and assumptions about the stellar populations and dust distributions in galaxies. The power of this method is most readily achieved, and the limitations best understood, when it can be combined with spectroscopic measurements of low-ionization atomic emission lines and small grains in individual galaxies in deep and wide area surveys.

\clearpage

\section*{Radio Probes of Star Formation and Black Hole Accretion}
\vspace{-0.1in}

\begin{wrapfigure}{r}{0.55\textwidth}
  \begin{center}
  \vspace{-0.45in}
\includegraphics[width=7.3cm]{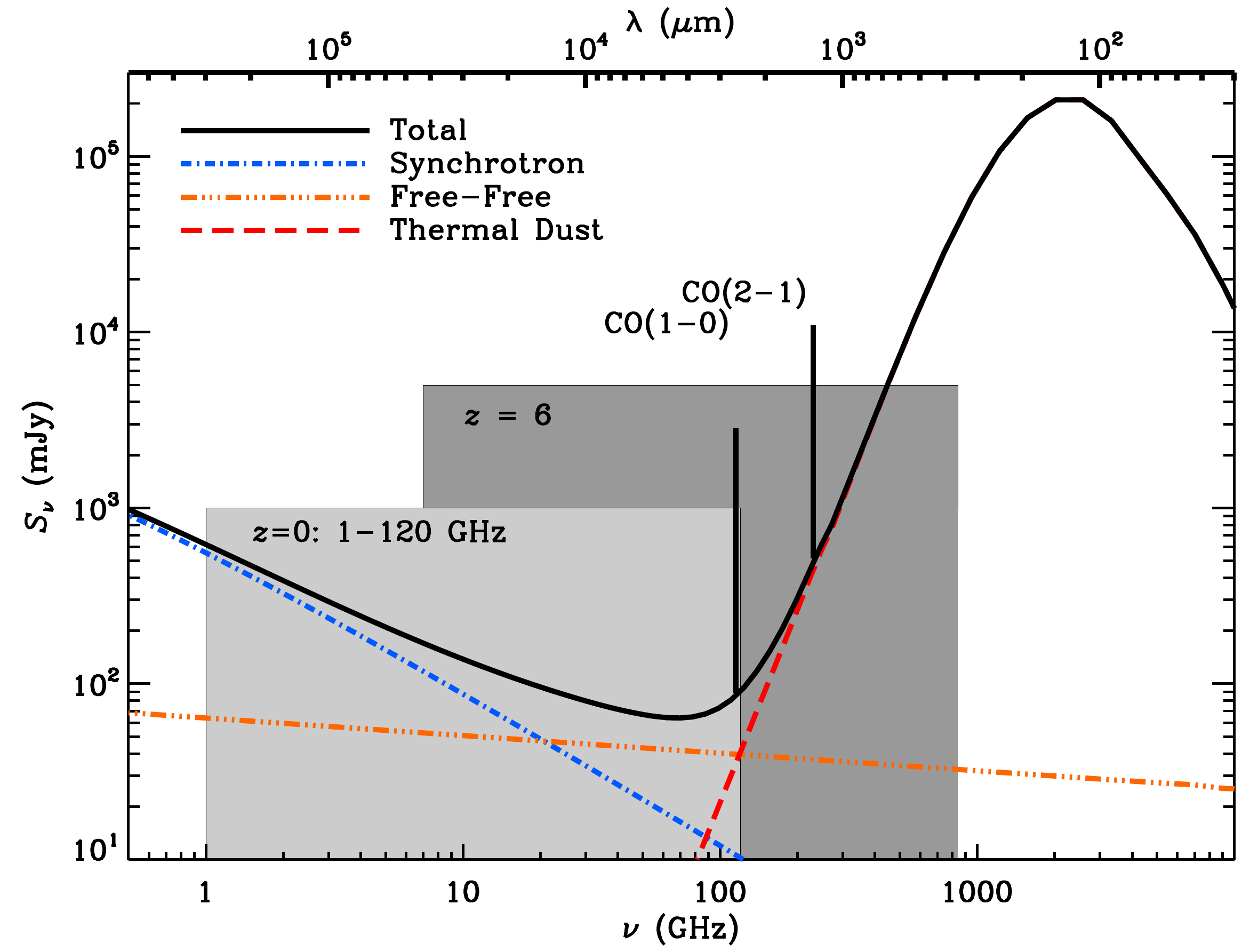}
  \end{center}
  \vspace{-0.32in}
  \caption{{\small 
 A model radio-to-infrared spectrum for a star-forming galaxy, showing that observations at frequencies spanning 1--120$\,$GHz provide direct access to free-free emission, whose contribution increases with frequency, and thus with the rest-frame emission from high-$z$ galaxies (highlighted for $z\approx6$).
 }
\label{fig:sed}}
  \vspace{-0.1in}
\end{wrapfigure}  
Given the complementary nature of the (extinction-free) diagnostics obtained in the radio with those obtained in the infrared,  joint investigations provide a powerful way to accurately distinguish and characterize star formation and black hole accretion energetics in the high redshift Universe. \\
\vspace{-0.1in}

\noindent{\bf Deep Continuum Surveys} Radio continuum emission from galaxies covering about 0.25--$30\,$cm (1--120$\,$GHz) is powered by a mix of physical emission processes, each providing independent information on the star formation and interstellar medium properties of galaxies (Figure \ref{fig:sed}).  
At frequencies $\lesssim\,$10\,GHz, radio continuum emission from star-forming galaxies is dominated by non-thermal synchrotron radiation arising from cosmic-ray (CR) electrons accelerated by supernova remnants. 
The other dominant process powering the radio continuum is free-free emission arising from ionized gas in H{\sc ii} regions, which is directly proportional to the production rate of ionizing (Lyman continuum) photons and is optically-thin at radio frequencies. 
Globally, free-free emission begins to dominate the total radio emission in normal star-forming galaxies at $\gtrsim\,$30\,GHz (e.g., Condon et al. 1992), with 30\,GHz free-free fractions measured as high as 80\% in nearby starbursts (Peel et al. 2011).   

From the well-established FIR-radio correlation (e.g., Helou et al. 1985; Condon 1992), we know that non-thermal synchrotron emission is a valuable tracer of the total amount of star formation in galaxies, unbiased by dust, and differences between the radio and far-infrared emission from high-$z$ galaxies is a powerful means to identify AGN (e.g., Yun et al. 2001).  
However, cosmic ray electrons lose energy through inverse Compton scattering off photons from the cosmic microwave background (CMB). This loss of energy scales as $(1+z)^{4}$ so that non-thermal radio emission from galaxies should become severally suppressed with increasing redshift.
This expectation is consistent with a recent 10\,GHz deep field observation for a sample of $z\gtrsim 1$ star-forming galaxies in the GOODS-N field finding that rest-frame 20\,GHz observations appear to be dominated by free-free emission (Murphy et al. 2017). Therefore, the free-free radio continuum (rest-frame $>30\,$GHz for galaxies at $z\gtrsim2$) is ideal for estimates of star formation activity at high redshift, unbiased by dust (Murphy 2009), making it an ideal complement to far-infrared based SFR estimates. \\
\vspace{-0.1in}

\noindent{\bf High-Resolution Investigations of Individual Sources} In addition, interferometric radio maps deliver high angular resolution to study the size and distribution of star formation and AGN activity in populations of high redshift galaxies.  
For instance, in moderate resolution (i.e., $\approx 1^{\prime\prime}$) maps, angular sizes in the radio can be used to measure SFR densities for entire galaxies, that can be used to identify starbursts (e.g., Murray et al. 2005).  With higher resolution ($\lesssim\,$0.01--0.1$^{\prime\prime}$) and enough sensitivity, the distribution of star formation can be mapped on sub-kpc scales in individual high-redshift galaxies, as well as the molecular gas distributions and dynamics, providing complementary information on star formation and AGN fueling (see White Papers by E.~Murphy et al.~and C.~Carilli et al.).
Furthermore, nuclear emission from such high ($\sim0.01^{\prime\prime}$) imaging can be extracted and characterized independently, to distinguish between disk star-formation and nuclear accretion energetics (e.g., through ancillary spectral diagnostics and/or brightness temperature arguments; Condon et al. 1991), and this provides an extremely powerful means to measure the co-evolution of star formation and black-hole accretion on a galaxy-by-galaxy basis.

\vspace{-0.2in}

\section*{Astro2020: Necessity for sensitive thermal IR and radio facilities}
\vspace{-0.1in}

\begin{figure}[!b]
\centering
\vspace{-0.2in}
\includegraphics[width=15cm]{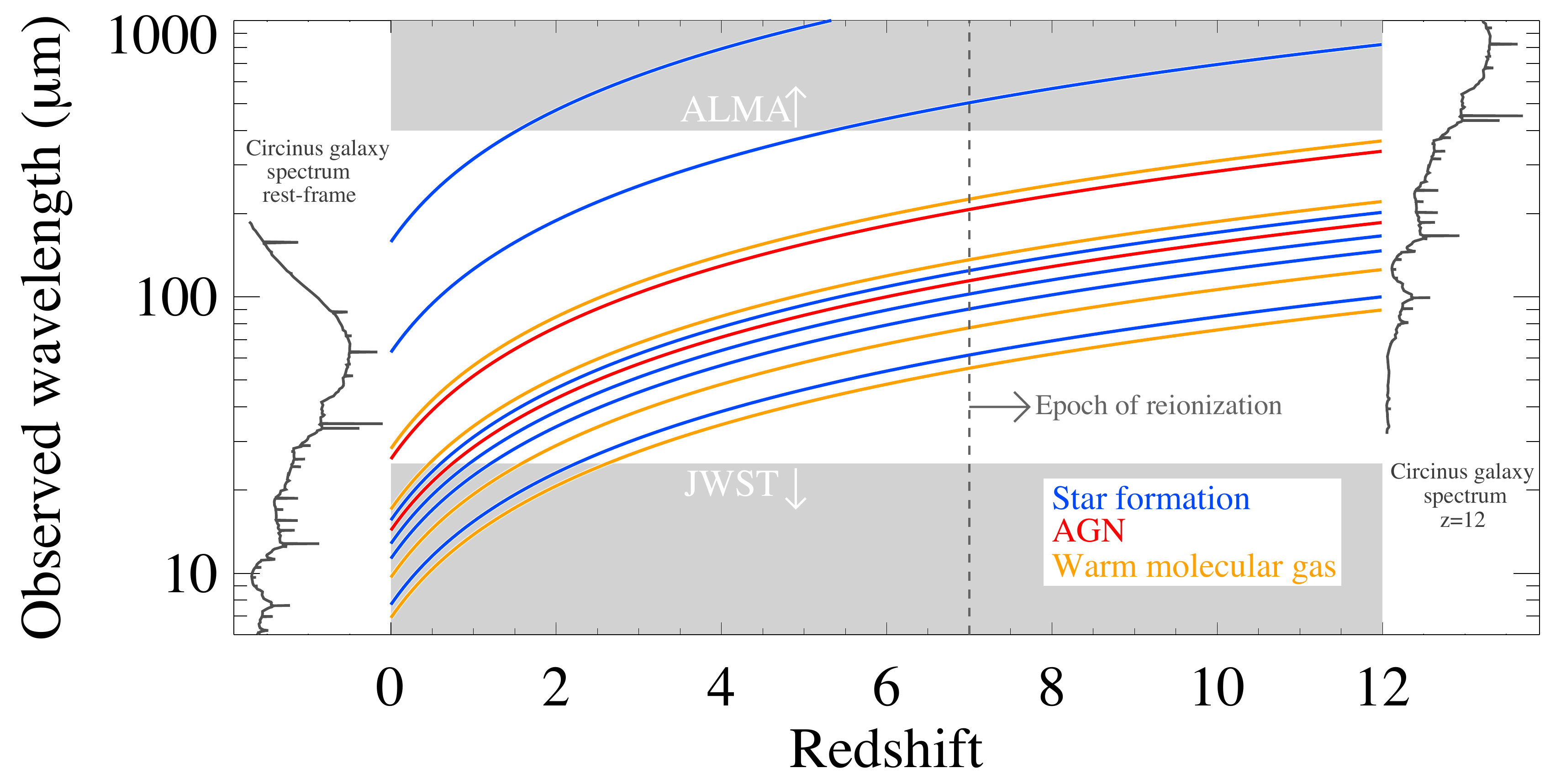}
\vspace{-0.2in}
\caption{\small There exists an order of magnitude gap in wavelength coverage between {\it JWST} and ALMA, where many powerful IR diagnostic lines fall at high redshift. We have yet to build a infrared space telescope with the sensitivity to detect most of these lines outside the local Universe.
\label{fig:linesvz}}
\end{figure}

In the coming decades, a definitive answer to the outstanding question of how galaxies and their supermassive black holes coevolve will require novel thermal infrared and radio facilities in space and on the ground.
 A future, cold infrared telescope in space, such as the {\it Origins Space Telescope}, is necessary to bridge the large wavelength gap between {\it JWST} and ALMA and access the key redshifted rest-frame mid- and far-infrared star formation and AGN diagnostic features (Figure 3). With $1000\times$ the sensitivity of previous space telescopes in the far-IR, {\it Origins} will spectroscopically map large areas of the sky (tens of deg$^2$), in order to build up large samples of galaxies across a wide variety of environments and characterize rare objects. Smaller, yet still capable, a probe-class cryogenic far-IR mission or ESA/JAXA's SPICA mission, would offer lower-cost alternatives and a first step toward addressing some of the questions motivated above.
 A ground-based radio interferometer, such as the ngVLA, would have the spatial resolution and sensitivity to separate the star formation and black hole accretion in individual galaxies, while also undertaking large-area surveys. 
 Together these new facilities will measure the SFRs and BHARs in individual galaxies from $z\,{=}\,0$--8, in order to determine how galaxy and supermassive black holes evolve together with cosmic time, and across the cosmic web.

\pagebreak
\noindent{\large {\bf References} }

\noindent Aird, J., Coil, A.~L., Georgakakis, A., et al.\ 2015, MNRAS, 451, 1892 \\
Alexander, D.~M., \& Hickox, R.~C.\ 2012, New Astronomy Reviews, 56, 93 \\
Armus, L., Charmandaris, V., Bernard-Salas, J., et al.\ 2007, ApJ, 656, 148 \\
Armus, L., Bernard-Salas, J., Spoon, H.~W.~W., et al.\ 2006, ApJ, 640, 204 \\ 
Bernard Salas, J., Pottasch, S.~R., Beintema, D.~A., \& Wesselius, P.~R.\ 2001, A\&A, 367, 949 \\
Brandl, B. R., et al. 2006, ApJ, 653, 1129 \\
Casey, C.~M., Zavala, J.~A., Spilker, J., et al.\ 2018, ApJ, 862, 77 \\
Condon, J.~J.\ 1992, ARA\&A, 30, 575\\
Condon, J.~J., et al.,\ 1991, ApJ, 378, 65\\
Dasyra, K.~M., Ho, L.~C., Armus, L., et al.\ 2008, ApJL, 674, L9 \\
Delvecchio, I., Gruppioni, C., Pozzi, F., et al.\ 2014, MNRAS, 439, 2736 \\
Devost, D.\ 2007, Bulletin of the American Astronomical Society, 39, 112.09 \\
Genzel, R., et al. 1998, ApJ, 498, 579 \\
Genzel, R., \& Cesarsky, C.~J.\ 2000, ARA\&A, 38, 761 \\
Gruppioni, C., Berta, S., Spinoglio, L., et al.\ 2016, MNRAS, 458, 4297 \\
Helou et al. 1985, ApJ, 298, L7\\
Hickox, R.~C., \& Alexander, D.~M.\ 2018, ARA\&A, 56, 625 \\
Ho, L.~C., \& Keto, E.\ 2007, ApJ, 658, 314 \\
Houck, J. R., et al. 2005, ApJ, 622, L105\\
Inami, H., Armus, L., Charmandaris, V., et al.\ 2013, ApJ, 777, 156\\
Kennicutt, R. C.~1998, ARA\&A, 36, 189 \\
Kirkpatrick, A., Pope, A., et al.\ 2015, ApJ, 814, 9 \\
Kormendy, J., \& Ho, L.~C.\ 2013, ARA\&A, 51, 511 \\
Laurent, O., Mirabel, I.~F., Charmandaris, V., et al.\ 2000, A\&A, 359, 887 \\
Lutz, D., Sturm, E., Genzel, R., et al.\ 2003, A\&A, 409, 867 \\
Madau P. \& Dickinson M., 2014, ARA\&A, 52, 415 \\
Murphy E. J., 2009, ApJ, 706, 482\\
Murphy, E.~J., Condon, J.~J., Schinnerer, E., et al.\ 2011, ApJ, 737, 67 \\
Murphy, E.~J., et al.\ 2017, ApJ, 839, 35\\ 
Murray, N., et al., 2005, ApJ, 618, 569\\
Peel M., et al. 2011, MNRAS, 416, L99\\
Petric, A.~O., Armus, L., Howell, J., et al.\ 2011, ApJ, 730, 28 \\
Pope, A., Chary, R.-R., Alexander, D.~M., et al.\ 2008, ApJ, 675, 1171 \\
Pottasch, S.~R., Beintema, D.~A., Bernard Salas, J., \& Feibelman, W.~A.\ 2001, A\&A, 380, 684 \\
Riechers, D.~A., Pope, A., Daddi, E., et al.\ 2014, ApJ, 786, 31 \\
Smith, J. D. T., et al. 2007, ApJ, 656, 770 \\
Smith, J.~D.~T., Dale, D.~A., Armus, L., et al.\ 2004, ApJS, 154, 199 \\
Veilleux, S., Rupke, D.~S.~N., Kim, D.-C., et al.\ 2009, ApJS, 182, 628 \\
Vito, F., Brandt, W.~N., Yang, G., et al.\ 2018, MNRAS, 473, 2378 \\
Yan, L., Chary, R., Armus, L., et al.\ 2005, ApJ, 628, 604 \\
Yun, M.~S., et al.\ 2001, ApJ, 554, 803 \\

\end{document}